%% file: 0_main.tex
\def\year{2022}\relax
\documentclass[letterpaper]{article} 
\usepackage{aaai22}  
\usepackage{times}  
\usepackage{helvet}  
\usepackage{courier}  
\usepackage[hyphens]{url}  
\usepackage{graphicx} 
\usepackage{natbib}  
\usepackage{caption} 
\DeclareCaptionStyle{ruled}{labelfont=normalfont,labelsep=colon,strut=off} 
\usepackage[title]{appendix}
\usepackage{booktabs}
\usepackage{multirow}
\usepackage{algorithmic}
\usepackage{enumitem}
\usepackage{makecell}
\usepackage{graphicx}
\usepackage{caption}
\usepackage{subfigure}
\usepackage[linesnumbered, ruled, vlined]{algorithm2e}
\usepackage{textcomp}
\usepackage{listings}
\usepackage{xcolor}
\input{0_math_def}

\begin{document}
\newcommand\yang[1]{\textcolor{magenta}{[Hongxia:  #1 ]}}
\title{Dynamic Sequential Graph Learning for Click-Through Rate Prediction}




\author {
    Yunfei Chu,
    Xiaofu Chang,
    Kunyang Jia, 
    Jingzhen Zhou,
    Hongxia Yang
}
\affiliations {
    Damo Academy, Alibaba Group\\
    {fay.cyf}@alibaba-inc.com, changxiaofu123@163.com, kunyang.jky@alibaba-inc.com,\\ jingren.zhou@alibaba-inc.com, yang.yhx@alibaba-inc.com
}


\renewcommand{\algorithmicrequire}{ \textbf{Input:}} 
\renewcommand{\algorithmicensure}{ \textbf{Output:}}





\maketitle

\newcommand\chu[1]{\textcolor{blue}{[yunfei TODO: #1]}}

\begin{abstract}
\input{0_abs.tex}
\end{abstract}

\section{Introduction}
\input{1_intro.tex}

\section{Related Work}
\input{2_rel.tex}
\section{Dynamic Sequential Graph Learning}
\input{3_method.tex}

\section{Experiments}
\input{4_experiment.tex}


\section{Conclusion and future work}

\input{5_con.tex}


\clearpage
\bibliography{0_dsgl.bib}

\end{document}

%% file: 0_math_def.tex
\usepackage{amsmath,amsfonts,bm}

\def\eqref#1{equation~\ref{#1}}

\def\1{\bm{1}}

\def\rve{{\mathbf{e}}}
\def\rvf{{\mathbf{f}}}

\def\rvh{{\mathbf{h}}}

\def\rvx{{\mathbf{x}}}

\def\mK{{\bm{K}}}

\def\mQ{{\bm{Q}}}

\def\mV{{\bm{V}}}
\def\mW{{\bm{W}}}

\def\gD{{\mathcal{D}}}
\def\gE{{\mathcal{E}}}
\def\gF{{\mathcal{F}}}
\def\gG{{\mathcal{G}}}
\def\gH{{\mathcal{H}}}

\def\gL{{\mathcal{L}}}

\def\gN{{\mathcal{N}}}

\def\gS{{\mathcal{S}}}

\def\sR{{\mathbb{R}}}



%% file: 0_abs.tex
Click-through rate prediction plays an important role in the field of recommender system and many other applications. 
Existing methods mainly extract user interests from user historical behaviors. However, behavioral sequences only contain users' directly interacted items, which are limited by the system's exposure, thus they are often not rich enough to reflect all the potential interests. 
In this paper, we propose a novel method, named Dynamic Sequential Graph Learning (DSGL), to enhance users or items' representations by utilizing collaborative information from the local sub-graphs associated with users or items. Specifically, we design the Dynamic Sequential Graph (DSG), i.e., a lightweight ego subgraph with timestamps induced from historical interactions. At every scoring moment, we construct DSGs for the target user and the candidate item respectively.
Based on the DSGs, we perform graph convolutional operations iteratively in a bottom-up manner to obtain the final representations of the target user and the candidate item. As for the graph convolution, we design a Time-aware Sequential Encoding Layer that leverages the interaction time information as well as temporal dependencies to learn evolutionary user and item dynamics. Besides, we propose a Target-Preference Dual Attention Layer, composed of a preference-aware attention module and a target-aware attention module, to automatically search for parts of behaviors that are relevant to the target and alleviate the noise from unreliable neighbors. Results on real-world CTR prediction benchmarks demonstrate the improvements brought by DSGL.

%% file: 1_intro.tex
Click-Through Rate (CTR) prediction is critical in many applications such as recommendation, online advertising and web search, and the main goal is to estimate the likelihood of a user
clicking at an item~\cite{zhou2019deep}. Since accurate CTR prediction benefits both business effectiveness and user experience, the topic has drawn the attention of both academic and industry communities.

On many online platforms, users interact with items in chronological order, forming the historical interaction sequences.
Motivated by deep learning’s expressive power in modeling sequential data, some sequential methods~\cite{hidasi2015session,wang2017topological,zhou2018atrank,tang2018personalized,guo2019dynamic,li2020deep} utilize the recurrent neural networks or the self-attention networks to model the update of user interest through the historical behaviors, and have gained impressive success in CTR prediction.
Despite the progress, the above methods only focus on mining the associations between the candidate item and the target user's historical behaviors, suffering from some limitations.  
On one hand, the user behavior sequences are limited by the recommender system’s exposure. Relying only on the one-hop collaborative neighbors of a user is hard to predict his/her emerging or potential interests. On the other hand, when the interactions are sparse, especially for inactive users whose sequences are short with long time intervals, it is hard to learn high-quality representations.

\begin{figure}
    \centering
    \includegraphics[width=1.0\linewidth]{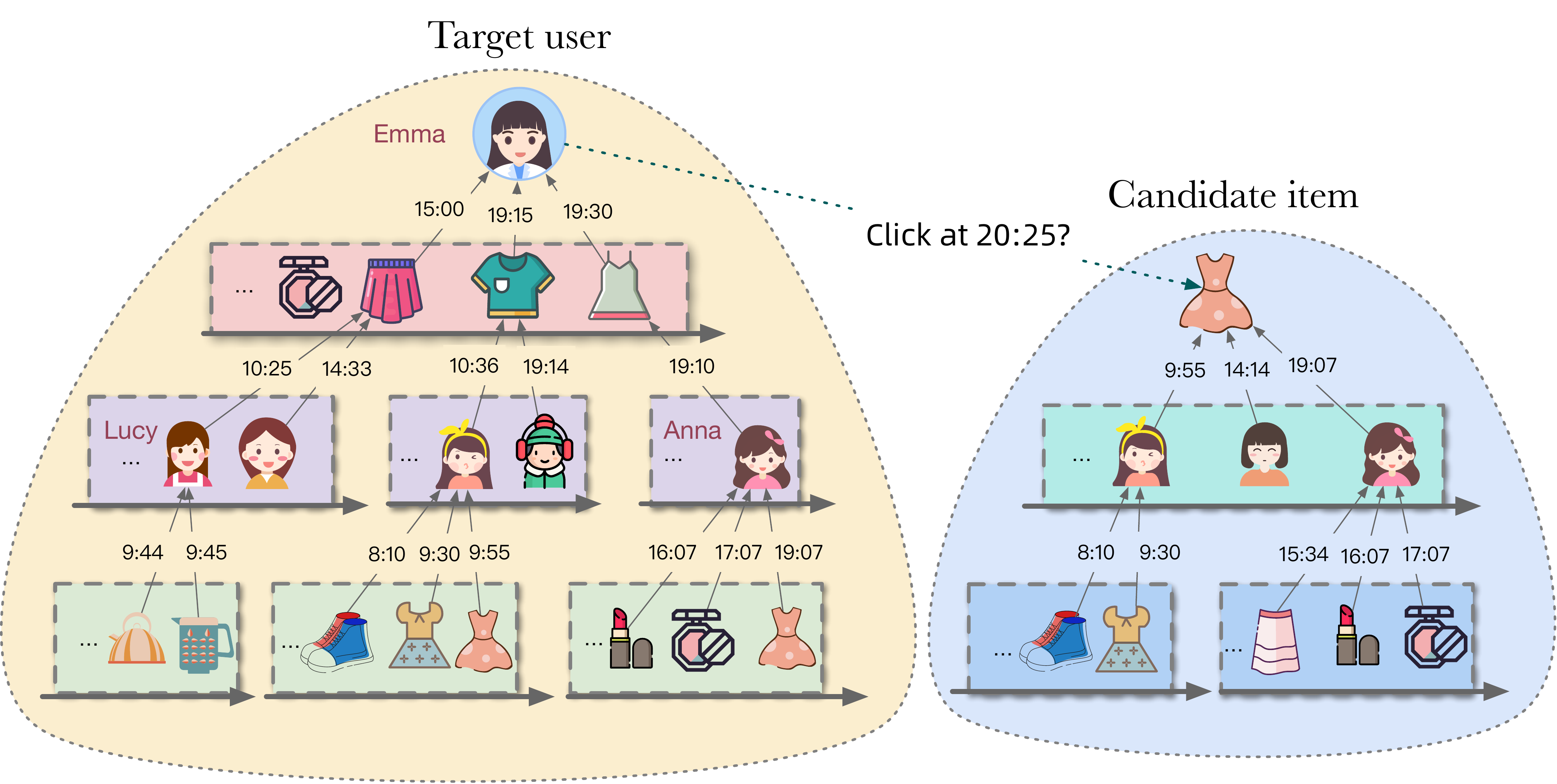}
    \caption{Illustration of Dynamic Sequential Graph for CTR prediction. DSG is a lightweight heterogeneous time-evolving ego graph combining the multi-hop connectivity in graphs and the temporal dependency in sequences. In CTR prediction, we construct DSGs for the target user and the candidate item at every scoring moment, respectively.}
    \label{fig:DSG_rec}
    \vspace{-6pt}
\end{figure}

To tackle these challenges, we propose a novel method called Dynamic Sequential Graph Learning (DSGL), which enriches the collaborative information explicitly by constructing dynamic sequential graphs and models the sequential evolution of the multi-hop collaborative neighbors. 
Specifically, we propose a novel lightweight heterogeneous time-evolving ego graph, namely Dynamic Sequential Graph (DSG), to capture the dynamics of both the target user and the candidate item from their respective multi-hop collaborative neighbors at every scoring moment, as illustrated in Figure~\ref{fig:DSG_rec}. 
Then, we perform graph convolution bottom-up iteratively on the DSGs, i.e., to learn the new representation of a node by aggregating the embeddings of the neighbors in its historical behavior sequence.
The graph convolution contains two main components: A Time-Aware Sequence Encoding Layer and A Target-Preference Dual Attention Layer.
The Time-Aware Sequence Encoding Layer leverages the sequential dependency and time decay information explicitly in the behavior sequence to capture the evolutionary dynamics. 
The Target-Preference Dual Attention Layer is composed of two module: An preference-aware attention mechanism that finds representative behaviors that are similar to the central node's preference, and a target-aware attention mechanism that searches for parts of behaviors that are relevant to the target node.   
Finally, inspired by~\cite{he2020lightgcn}, DSGL combines the representations learned at different graph convolution layers with a weighted sum to obtain the final embedding for prediction.

Our main contributions can be summarized as follows:
\begin{itemize}[leftmargin=*]
\item We construct lightweight dynamic sequential graphs for CTR prediction. To the best of our knowledge, this work is the first dynamic-graph-based CTR prediction method.
\item  We propose a model DSGL that performs graph convolution on DSGs. The graph convolution operation consists of a Time-Aware Sequence Encoding Layer as well as a Target-Preference Dual Attention Layer to capture the evolutionary dynamics for nodes and alleviate the noise brought by the unreliable neighbors.
\item We conduct extensive experiments on real-world CTR prediction benchmarks. Experimental results demonstrate the effectiveness of our model over strong and state-of-the-art baselines.
\end{itemize}

%% file: 2_rel.tex
We discuss two lines of researchs that are relevant to our work: 1) deep models for CTR prediction, and 2) recent progress of graph neural network-based methods developed for recommendation.
\subsection{Deep Models for CTR Prediction}
The prediction of Click-Through Rate (CTR) plays an important role in many applications, ranging from web search, personalized recommendation and online advertising. 
 In recent years, deep learning based CTR prediction models have achieved remarkable success in feature interaction modeling~\cite{guo2021dual,lyu2020deep}. Wide\&Deep~\cite{cheng2016wide} and DeepFM~\cite{guo2017deepfm} combines the advantage of shallow model and non-linear deep model to learn low-order and high-order feature interactions simultaneously. PNN~\cite{qu2016product} introduces a product layer to capture high-order feature interactions between inter-field categories. However, these methods cannot capture the interest behind data clearly.

Nowadays, extracting user interest from the historical behavior for better CTR prediction has attracted increasing attention. 
Recurrent Neural Networks (RNN) are commonly adopted due to their power in modeling sequence~\cite{hidasi2015session,wu2017recurrent,hidasi2018recurrent}. Among them, GRU4Rec~\cite{hidasi2015session} is the first RNN-based models that uses Gated Recurrent Units (GRU)~\cite{cho2014learning} to capture the dependencies in users' historical behavior sequence.  
Besides RNN, Convolutional Neural Networks (CNN) have also been applied to learn sequential patterns using convolutional filter~\cite{tang2018personalized}.
In order to selectively utilize information from interactions that are truly relevant to the next interaction prediction, attention-based models are increasingly employed~\cite{ying2018sequential,zhou2018deep,zhou2018atrank,zhou2019deep,feng2019deep}.
For example, DIN~\cite{zhou2018deep} and DIEN~\cite{zhou2019deep} designs an attention mechanism to model the users’ interests from historical behaviors w.r.t. the target item. 
Above methods only model the user behavior with the evolution of items ignored. Motivated by it, TIEN~\cite{li2020deep} deals with the item behavior by proposing an attention mechanism to achieve robust personalized item dynamics. 

Despite great success has been made by the above CTR prediction methods, they cannot explore the user-item interactions to the fullest extent, since they only consider the one-hop behavior. 
We are going to solve the problem in this paper by incorporating and exploiting the graph learning to enrich the collaborative information.

\subsection{Graph Neural Network for Recommendation}
In the last few years, Graph Neural Networks (GNN) have seen a great surge of interest with promising methods~\cite{graphsage,velivckovic2017graph}. The effectiveness of GNNs is also proved on recommendation problems.
Historically, two classes of GNN-based recommendation have been developed.

The first class considers the user-item interactions as a static bipartite graph and adapts GCN to the user-item graph, capturing Collaborative Filtering (CF) signals in high-hop neighbors for recommendation~\cite{berg2017graph,ying2018graph,wang2019neural,he2020lightgcn}. However, these methods compress the dynamic time-evolving interactions into a static snapshot, disregarding the time-dependent structure, the exact timestamps (or time intervals), and emerging interactions. Thus, they fail to capture the dynamics in users and items.

The second class models the interactions as dynamic graphs. One branch of dynamic-graph-based methods~\cite{goyal2018dyngem,pareja2020evolvegcn} utilize a sequence of discrete snapshots to model the time-evolving interactions, but they cannot capture the fine-grained temporal and structural information, thus failing to capture the real-time interests. Although some methods~\cite{dai2016deepcoev,jodie} model dynamic graphs in continuous time and update dynamic node embeddings given each interaction, they fail to capture the higher-order temporal neighborhood structures explicitly. 
Another branch of such methods~\cite{wu2019session,yu2020tagnn,song2019session,chen2020handling} model the temporal dependency of interacted items in behavior sequences as session graphs, and then adopt GNN to capture the complex transitions of items. Although these methods utilize time-dependent structure to model dynamic user interests, similar to sequential model in CTR prediction, they cannot capture high-order connectivity, thus confining the performance. 
 
Our work differs from above methods in that DSGL constructs lightweight ego subgraphs for the target user and candidate item at every ranking moment and combines the advantages of graphs and sequences to capture high-order connectivity and temporal dependency simultaneously.

%% file: 3_method.tex
\begin{figure}
    \centering
    \includegraphics[width=\linewidth]{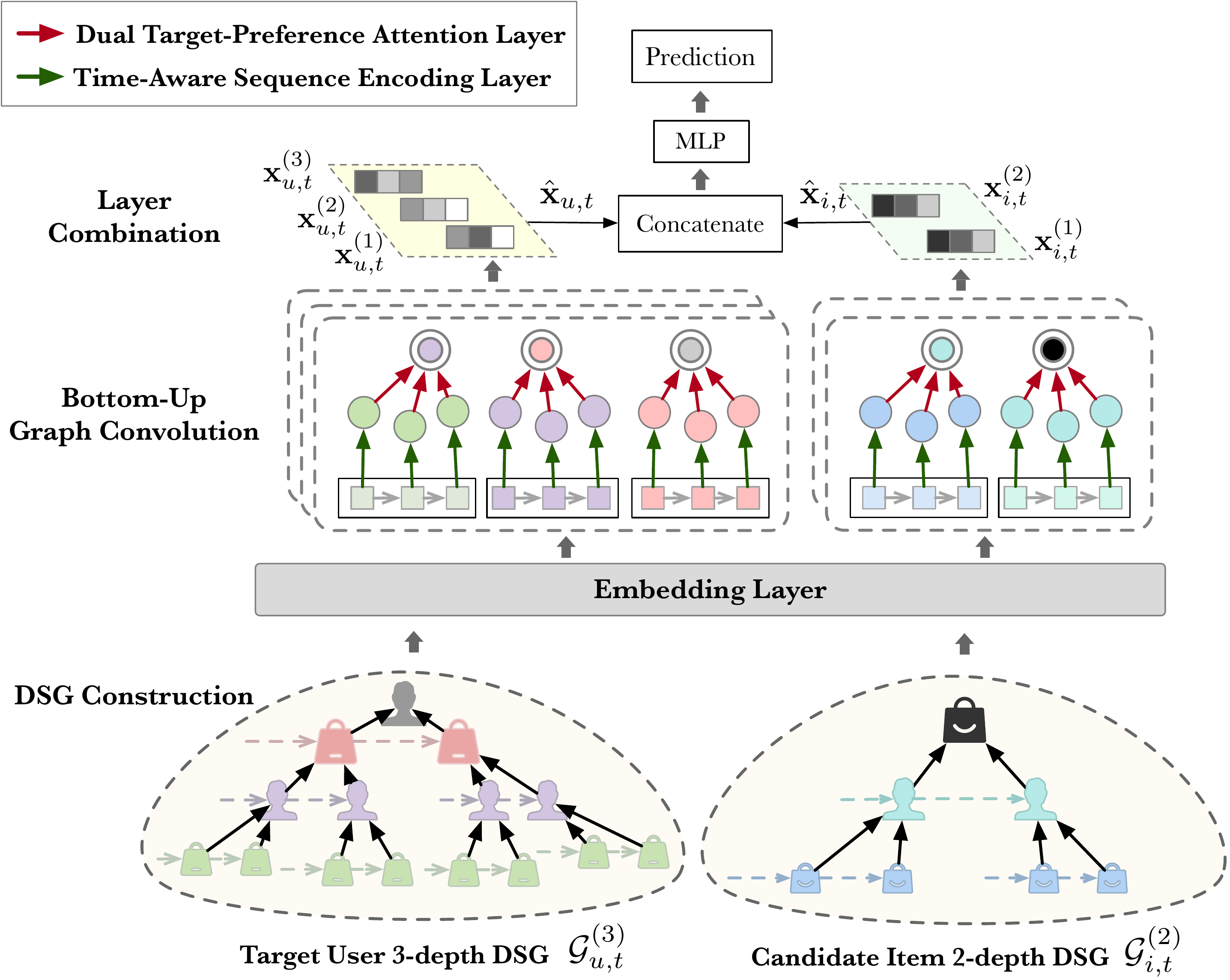}
    \caption{Framework of the proposed DSGL method. At every scoring moment, DSGs are constructed for the target user $u$ (left) and the candidate item $i$ (right) respectively. Their representations are refined with multiple bottom-up graph convolutions, each of which consists of a Time-Aware Sequence Encoding Layer and a Target-Preference Dual Attention Layer. DSGL gets the final representations via layer combination followed by an MLP-based prediction layer.}
    \label{fig:DSGL}
\end{figure}
The basic idea of DSGL is to perform graph convolution iteratively on the DSGs to enrich the collaborative information explicitly. 
In this section, we elaborate the design of Dynamic Sequential Graph Learning (DSGL).
The methods consists of two main components:
(1) Dynamic Sequential Graph Construction that establishes multi-hop collaborative connections to expand user and item behaviors.
(2) Bottom-Up Graph Convolution that refines the node embedding by aggregating the embeddings of the neighbors in the behavior sequence in a bottom-up manner.
Specifically, the Bottom-Up Graph Convolution consists of two layers: the Time-Aware Sequence Encoding Layer that encodes the behavior sequence with time information and temporal dependency captured
and the Target-Preference Dual Attention Layer that activates the related behavior in the sequence to eliminate noisy information.
Besides the above components, we also propose an embedding layer that initializes user, item, and time embeddings, a layer combination module that combines the embeddings of multiple layers to get final representations, and a prediction layer that outputs the prediction score.
The complete framework is demonstrated in Figure~\ref{fig:DSGL}.

\subsection{Dynamic Sequential Graph Construction}
The way we define a graph is vital for model performance. One can construct a heavy static graph containing all the users and items. However, such treatment neglect time-dependent structure and the exact timestamps, which is critical to capture potential or emerging preference.
Thus, we design a lightweight heterogeneous time-evolving ego graph, namely DSG, which combines the multi-hop connectivity in graphs and the fine-grained temporal dependency in sequences.

The DSGs are induced of historical user-item interaction edges $\gE=\left\{\left(u, i,t\right)\cup\left(i, u,t\right)\right\}$, each of which represents a user $u$ interacts with an item $i$ associated with an timestamp $t\in\sR^{+}$.
For each user-item pair to be scored, the corresponding DSG is constructed in a recursive way, and the process is as follows:
\begin{itemize}[leftmargin=*]
    \item For a user $u$ (or an item $i$) at time $t$, we define the $1$-depth DSG of user $u$ (or item $i$) at time $t$ as a set of directed interaction edges before time $t$ in chronological order, denoted by $\gG_{u,t}^{(1)}=\{(i,u,\tau)|\tau<t,(i,u,\tau)\in\gE\}$ (or $\gG_{i,t}^{(1)}=\{(u,i,\tau)|\tau<t,(u,i,\tau)\in\gE\}$).
    \item We define the ($k$+1)-depth DSG of user $u$ (or item $i$) at time $t$ as a set of $k$-depth DSGs that user $u$ (or item $i$) interacts in chronological order with its $1$-depth DSG, $\gG_{u,t}^{(k+1)}=\{\gG_{i,\tau}^{(k)}|\tau<t,(i,u,\tau)\in\gE\}\cup\gG_{u,t}^{(1)}$ (or $\gG_{i,t}^{(k+1)}=\{\gG_{u,\tau}^{(k)}|\tau<t,(u,i,\tau)\in\gE\}\cup\gG_{i,t}^{(1)}$).
\end{itemize}
\begin{figure}
    \centering
    \setlength\abovecaptionskip{2pt}
    \includegraphics[width=0.9\linewidth]{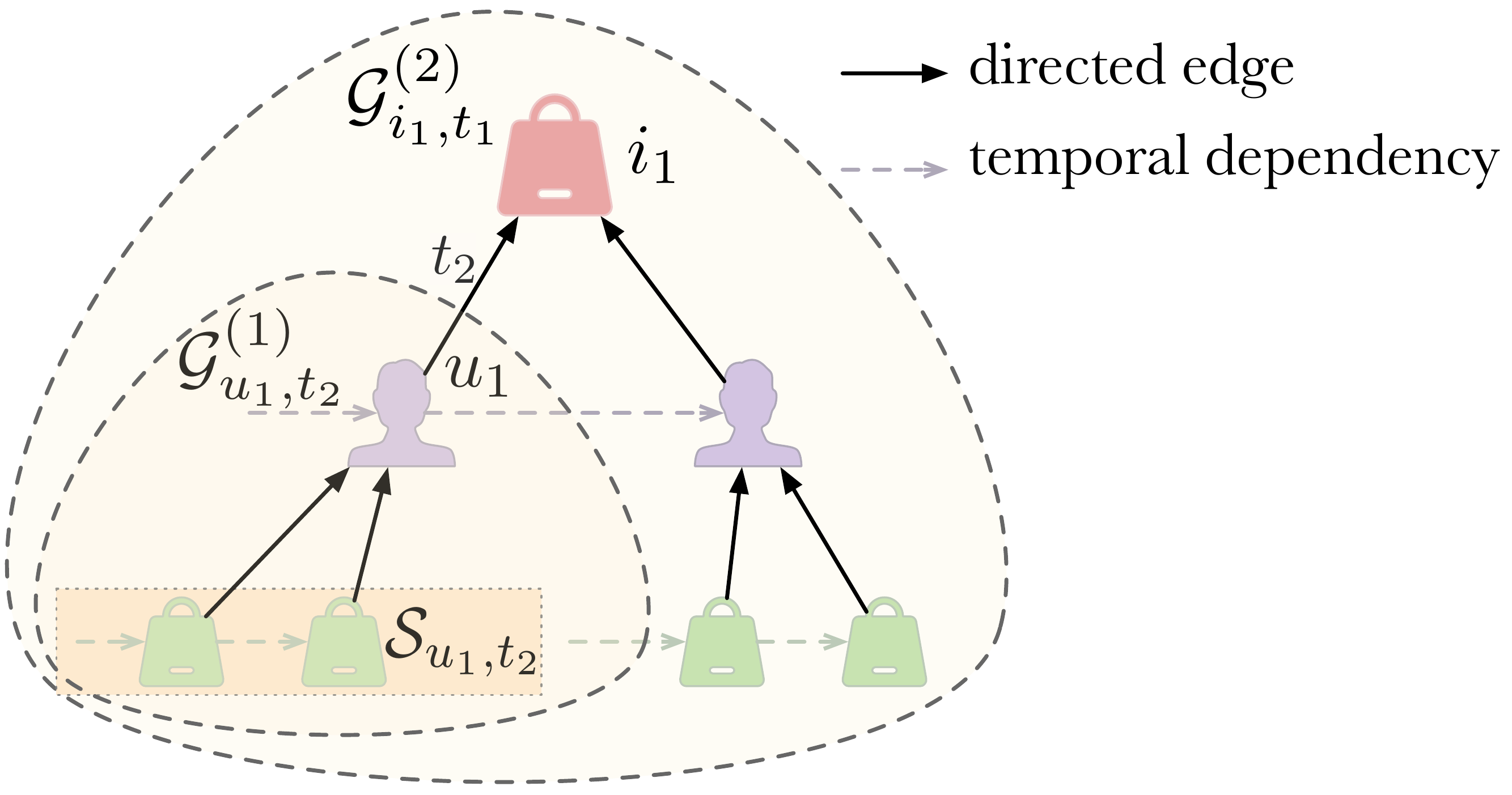}
    \caption{An example of an item's 2-depth dynamic sequential graph. DSG is constructed from bottom to top recursively. The edges are directed representing how messages are passed, and the direction can only be from the past node to the current node.}
    \label{fig:DSG}
\end{figure}
The construction process of DSG is illustrated as Figure~\ref{fig:DSG}. 
We define the historical behavior sequence of user $u$ (or item $i$) at time $t$ as a sequence of interacted items (or users) in chronological order, denoted by $\gS_{u,t}=\{(i,\tau)|\tau<t,(i,u,\tau)\in\gE\}$ (or $\gS_{i,t}=\{(u,\tau)|\tau<t,(u,i,\tau)\in\gE\}$).
In practice, we select the last $r$ interactions before time $t$ at each layer and control the number of layers to 2-3 layers to ensure DSG's lightweight from the perspective of efficiency. Note that the depth of the candidate item's DSG is set one less than that of the target user so that the structure of each item in the target user's sequence is the same as that of the candidate item.

\subsection{Embedding Layer}
There are two groups of inputs in the proposed DSGL: the target user's $k$-depth DSGs $\gG_{u,t}^k$ and the candidate item's ($k$-1)-depth DSGs $\gG_{i,t}^{k-1}$.  Features related to users can be user ID and user profile, e.g. age, gender, country, and so on. For items, features can be item ID and item attributes, such as category, brand, and statistical click-through rate. For DSGs, besides the node features, each interaction is associated with a timestamp.

For each field of discrete features, we represent it as an embedding matrix, and perform embedding lookups by feature ID to obtain low-dimensional embeddings of each discrete feature. By concatenating all fields of feature embeddings, we have the node embedding of items and users, denoted by $\rvf_{item}\in\sR^{d_i}$ and $\rvf_{user}\in\sR^{d_u}$.
As for the interaction timestamp in DSG, we compute the time intervals between the interaction time and its parent interaction time as time decays. For example, given a historical behavior sequence $\gS_{u,t}$ of user $u$ at the timestamp $t$, each interaction $(u,i,\tau)\in\gS_{u,t}$ corresponds to a time decay $\Delta_{(u,i,\tau)}=t-\tau$. Following~\cite{li2020deep}, we transform the continuous time decay values to discrete features by mapping them to a series of buckets with the ranges $[b^0, b^1), [b^1, b^2),\dots, [b^l, b^{l+1})$, where the base $b$ is a hyper-parameter. Then by performing the embedding lookup operation, the time decay embedding can be obtained, denoted by $\rvf_{time}\in\sR^{d_t}$.

\subsection{Time-Aware Sequence Encoding Layer }
The nodes at each layer of DSGs are in time order, which reflects the time-varying preference of users as well as the popularity evolution of items. Thus we perform sequence modeling as a part of graph convolution to capture the dynamics.

Previous works~\cite{hidasi2015session,li2020deep} usually apply Recurrent Neural Network (RNN) based model to the node feature in the historical sequence recurrently to refine the behavior embeddings. They preserve only the order of behaviors in a sequence, with the impacts of different time decays ignored. However, the time decay feature is critical in recommendation. On the one hand, the time information can reflect the drifting of user interests. Users may not be interested in the item that they interacted with far from the current anymore. On the other hand, the time information indicates the varying audience of items. 

Thus, we propose the Time-Aware Sequence Encoding Layer to capture the fine-grained time information explicitly. 
For each interaction $(u,i,t)$, we have the historical behavior sequence $\gS_{u,t}$ of user $u$ and $\gS_{i,t}$ of item $i$. For sequence $\gS_{u,t}$, by feeding each interacted item along with the time decay in the sequence into the embedding layer, the behavior embedding sequence is formed with the combined feature sequence, as $\{\rve_{i,\tau}|(i,\tau)\in\gS_{u,t}\}$, 
where $\rve_{i,\tau}=[\rvf_{{item}_i};\rvf_{{time}_\tau}]\in\sR^{d_{i}+d_{t}}$ is the embedding of item $i$ in the sequence. Similarly, for sequence $\gS_{i,t}$, we have the embedding sequence as $\{\rve_{u,\tau}|(u,\tau)\in\gS_{i,t}\}$, where $\rve_{u,\tau}=[\rvf_{{user}_u};\rvf_{{time}_\tau}]\in\sR^{d_{i}+d_{t}}$. 
We take the obtained embedding as the zero-layer of inputs, i.e., $\rvx_{u,t}^{(0)}=\rve_{u,t}$ and $\rvx_{i,t}^{(0)}=\rve_{i,t}$. For ease of notation, we will drop the superscript in the rest of the following two subsections.

In this layer, we infer the hidden state of each node in the behavior sequence step by step with the embeddings containing time information as inputs. The encoder can be LSTM~\cite{hochreiter1997long} or GRU~\cite{chung2014empirical}, whose gates can utilize time feature to control the information to be propagated with the time decay feature as part of input.
Given the behavior sequences $\gS_{u,t}$ and $\gS_{i,t}$, we represent $j$-th item's hidden states and inputs in the sequence $\gS_{u,t}$ as $\rvh_{{item}_j}$ and $\rvx_{{item}_j}$, and $j$-th user's hidden states and inputs in the sequence $\gS_{i,t}$ as $\rvh_{{user}_j}$ and $\rvx_{{user}_j}$. The forward formulas are 
\begin{equation}
    \begin{split}
    & \rvh_{{item}_j} = \gH_{item}(\rvh_{{item}_{j-1}},\rvx_{{item}_{j}});\\
    & \rvh_{{user}_j} = \gH_{user}(\rvh_{{user}_{j-1}},\rvx_{{user}_{j}}).\\
    \end{split}
\end{equation}
where $\gH_{user}(\cdot,\cdot)$ and $\gH_{item}(\cdot,\cdot)$ represent the encoding functions specific to user and item, respectively.

We obtain the corresponding hidden states sequence of historical behavior sequence $\gS_{u,t}$ and $\gS_{i,t}$ after the Time-Aware Sequence Encoding Layer. For ease of notation, the process can be represented as:
\begin{equation}
    \begin{split}
    & f_{\text{ENC}_{item}}(\{\rvx_{i,\tau}|(i,\tau)\in\gS_{u,t}\}) = \{\rvh_{i,\tau}|(i,\tau)\in\gS_{u,t}\}; \\
    & f_{\text{ENC}_{user}}(\{\rvx_{u,\tau}|(u,\tau)\in\gS_{i,t}\}) = \{\rvh_{u,\tau}|(u,\tau)\in\gS_{i,t}\}.
    \end{split}
\end{equation}

\subsection{Target-Preference Dual Attention Layer}
In practice, with the multi-hop connectivity in the graph, it is almost inevitable to introduce unreliable and noisy neighbors due to unintended interactions, fake similar users, and drifting interests or trends~\cite{zhou2018atrank,li2020deep}. To eliminate the noise, we propose a Target-Preference Dual Attention Layer that can perceive the core preferences (i.e.,the most representative behavior) meanwhile activate the most related behavior concerning the target. Thus, the attention mechanism considers the following two aspects.

\begin{figure}
    \centering
    \setlength\abovecaptionskip{2pt}  
    \includegraphics[width=1.0\linewidth]{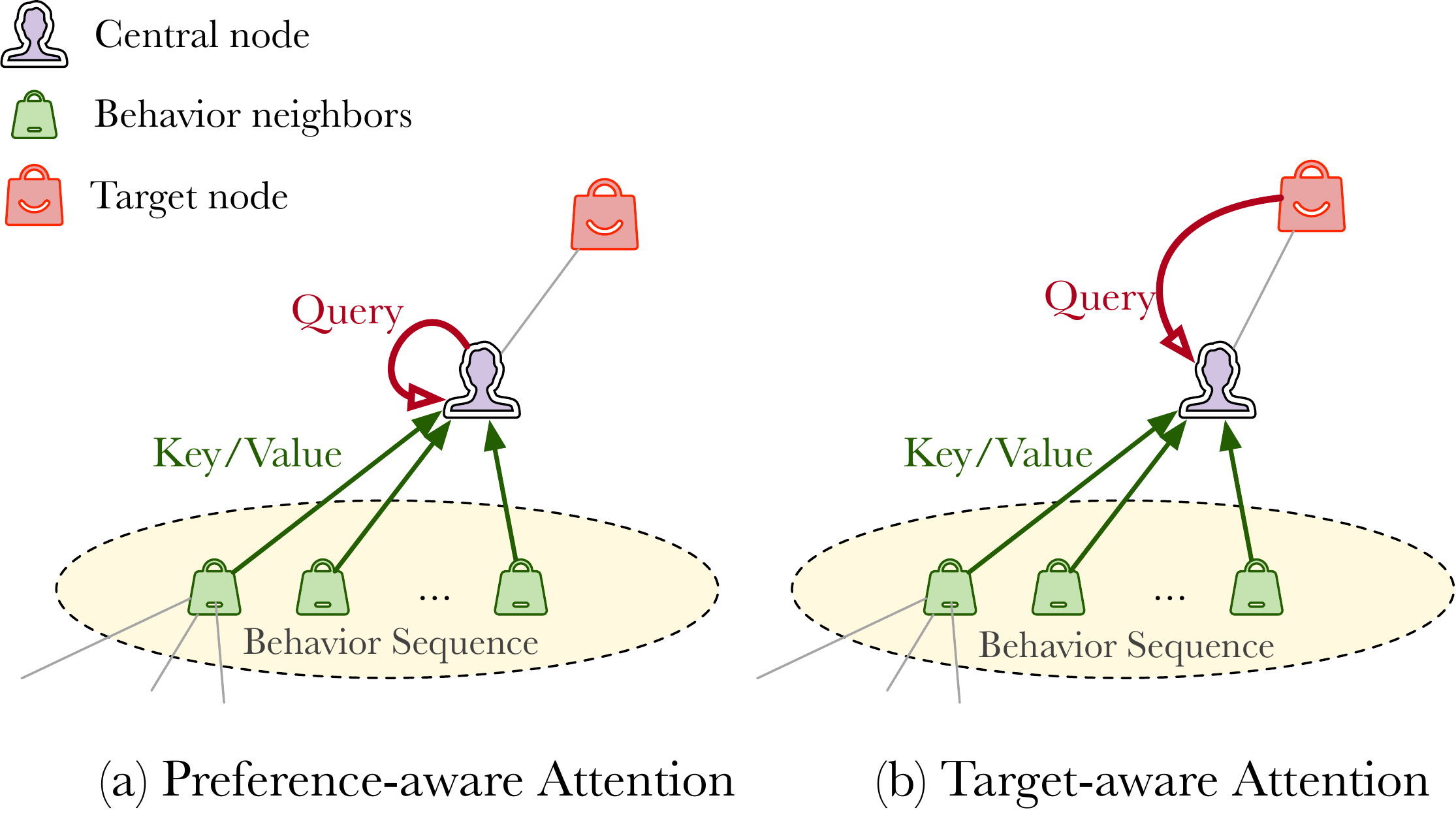}
    \caption{An illustration of the Target-Preference Dual Attention.}
    \setlength\belowcaptionskip{-2pt}  
    \label{fig:att}
    \vspace{-0.3cm}
\end{figure}

Firstly, we propose the preference-aware attention mechanism to perceive the core or main preferences of a user or an item from its behaviors. We computes the attention weights between the central node and its behavior neighbor nodes, which indicates the importance of each behavior neighbor to the central node. In this way, the central node is regarded as the attention query to activate the nodes most similar to 
it in its behavior sequence, as illustrated in Figure~\ref{fig:att}(a). Formally, given a user $u$ (or an item $i$) at time $t$ as the central node, we build the query of the preference-aware attention as:
\begin{equation}
    \mQ_{pa}=\rvx_{u,t}\quad\textit{or}\quad
    \mQ_{pa}=\rvx_{i,t}
\end{equation}

Secondly, we design the target-aware attention mechanism to soft-search parts of the central node's behavior sequences that are relevant to its target node. In DSGs, we take the upstream node of the central node as the target node, as illustrated in Figure~\ref{fig:att}(b). 
We explain this by the following example.
We take \emph{Anna} in Figure~\ref{fig:DSG_rec} as the central node. From \emph{Anna}'s behaviors, we find that she is interested in makeup products most of the time, and browses women's clothing occasionally. She is connected to Emma due to the co-click action on the \emph{green dress} (i.e., the upstream node). By taking the \emph{green dress} as the query, the behavior related to clothing can be activated and the preference about clothing can be further delivered upward, which can enrich the collaborative information about clothing for the target user \emph{Emma}.
For the target user and the candidate item, as they are the root nodes of their respective DSGs, we take them as the target nodes of each other to attend related behaviors like previous works~~\cite{zhou2018deep,zhou2019deep}. 
Formally, given a user $u$ (or an item $i$) at time $t$ as the central node, we build the query of the target-aware attention as:
\begin{equation}
    \mQ_{ta}=\rvx_{target(u,t)}\quad\textit{or}\quad
    \mQ_{ta}=\rvx_{target(i,t)}
\end{equation}
where $target{(*,t)}$ denotes the target node of the central node $*$ at time $t$. 

We obtain the Target-Preference Dual Attention Layer by summing up the results of preference-aware attention and the target-aware attention. In detail, we apply scaled dot-product attention as the attentive pooling method following~\cite{vaswani2017attention}, and the attention function is defined as
\begin{equation}
    \text{Attention}(\mQ,\mK,\mV)=\frac{softmax(\mQ\mK^T)}{\sqrt{d}}\mV
\end{equation}
where $\mQ$, $\mK$ and $\mV$ represent the \emph{Query}, \emph{Key} and \emph{Value}, respectively, and $d$ is the dimension of $\mK$ and $\mQ$. 
We take the behavior neighbors as the \emph{Key} and \emph{Value}.
Moreover, we adopts the multi-head attention~\cite{vaswani2017attention} to capture multiple interests or audience, defined as follows:
\begin{equation}
    \text{MultiHead}(\mQ,\mK,\mV)=[\text{head}_1;\text{head}_2;\dots;\text{head}_h]\mW_O
\end{equation}
\begin{equation}
    \text{head}_i=\text{Attention}(\mQ\mW_{Q_i},\mK\mW_{K_i},\mV\mW_{V_i})
\end{equation}
where weights $\mW_Q$, $\mW_K$, $\mW_V$ and $\mW_O$ are trained parameters.

Given the behavior hidden states sequence $\{\rvh_{i,\tau}|(i,\tau)\in\gS_{u,t}\}$ and $\{\rvh_{u,\tau}|(u,\tau)\in\gS_{i,t}\}$, we represents the attention process as:
\begin{equation}
    \begin{split}
    & \rvx_{{u,t}} = f_{\text{ATT}_{item}}(\{\rvh_{i,\tau}|(i,\tau)\in\gS_{u,t}\}); \\
    & \rvx_{{i,t}} = f_{\text{ATT}_{user}}(\{\rvh_{u,\tau}|(u,\tau)\in\gS_{i,t}\}). 
    \end{split}
\end{equation}

\subsection{Bottom-Up Graph Convolution and Layer Combination}
\begin{algorithm}
\caption{The algorithm of DSGL.} 
\centering
\label{alg:Framwork} 
\begin{algorithmic}[1]
\REQUIRE ~~\\ 
The training set $\gD=\{(u,i,t,y)\}$; Interaction set $\gE$; Depth $K$.\\
\ENSURE ~~ 
Network parameters $\Theta$.
\STATE Initialize input embedding $\rvf_{{user}}$ and  $\rvf_{{item}_i}$;\\ 
\FOR{$(u,i,t,y)\in\gD$}
\STATE Construct DSGs $\gG_{u,t}^{(K)}$, $\gG_{i,t}^{(K-1)}$ for user $u$ and item $i$;
\FOR{$(v,j,\tau)\in\gG_{u,t}^{(K-1)}\bigcup\gG_{i,t}^{(K-2)}$}
\STATE $\rvx_{v,\tau}^{(0)}\leftarrow{\rve_{v,\tau}}; \quad
\rvx_{j,\tau}^{(0)}\leftarrow{\rve_{j,\tau}}$;
\FOR{$k\leftarrow{1}$ to $K$}
\STATE $\rvx_{v,\tau}^{(k)}\leftarrow{f_{\text{AGG}_{item}}(\{\rvx_{j,\tau}^{(k-1)}|i\in\gS_{v,\tau}\})}$;
\STATE $\rvx_{j,\tau}^{(k)}\leftarrow{f_{\text{AGG}_{user}}(\{\rvx_{v,\tau}^{(k-1)}|i\in\gS_{j,\tau}\})}$;
\ENDFOR
\STATE $\hat{\rvx}_{{u,t}} \leftarrow \frac{1}{K}\sum_{k=1}^{K}\rvx_{{u,t}}^{(k)};
\hat{\rvx}_{{i,t}} \leftarrow \frac{1}{K-1}\sum_{k=1}^{K-1}\rvx_{{i,t}}^{(k)}$;
\STATE $\hat{y}_{u,i,t} \leftarrow \text{MLP}([\rve_{u,t};\rve_{i,t};\hat{\rvx}_{u,t};\hat{\rvx}_{i,t}])$;
\STATE Update the parameters $\Theta$ by optimizing Eq.\ref{eq_loss};
\ENDFOR
\ENDFOR
\end{algorithmic}
\end{algorithm}
The core idea of graph convolutions is to learning representation for nodes by performing convolution over their neighborhood. 
Note that we perform graph convolutions in a bottom-up manner to avoid the messages propagated from future.
The convolution computation for node $u$ at the $k$+1-th layer, which takes the input feature representation $\rvx_u^{(k)}$ and $\{\rvx_i^{(k)}|i\in\gN_{u}\}$ as input and outputs the induced representation $\rvx_u^{(k+1)}$ , can be abstracted as:
\begin{equation}
    \rvx_u^{(k+1)}={f_\text{AGG}}(\rvx_u^{(k)},\{\rvx_i^{(k)}|i\in\gN_{u}\})
\end{equation}
In DSGL, we stack the Time-Aware Sequence Encoding Layer and the Target-Preference Dual Attention, and the graph convolution operation can be represented as:
\begin{equation}
    \begin{split}
    & \rvx_{{u,t}}^{(k+1)} = f_{\text{ATT}_{item}}(f_{\text{ENC}_{item}}(\{\rvx_{i,t}^{(k)}|i\in\gS_{u,t}\})); \\
    & \rvx_{{i,t}}^{(k+1)} = f_{\text{ATT}_{user}}(f_{\text{ENC}_{user}}(\{\rvx_{u,t}^{(k)}|i\in\gS_{i,t}\})).
    \end{split}
\end{equation}

Different from traditional GCN models that use the last layer as the final node representation, inspired by ~\cite{he2020lightgcn}, we combine the embeddings obtained at each layer to form the final representation of a user (or an item):
\begin{equation}
\hat{\rvx}_{{u,t}} = \frac{1}{K_u}\sum_{k=1}^{K_u}\rvx_{{u,t}}^{(k)}; \quad
\hat{\rvx}_{{i,t}} = \frac{1}{K_i}\sum_{k=1}^{K_i}\rvx_{{i,t}}^{(k)},
\label{eq:lc}
\end{equation}
where $K_u$ and $K_i$ denotes the numbers of DSGL layers for user $u$ and item $i$, respectively.
\subsection{Model Prediction}
Given an interaction triplet $(u,i,t)$, with the corresponding DSGs of the target user and the candidate item, we can predict the possibility of the user interacting with the item as:
\begin{equation}
    \hat{y} = \gF(u,i,\gG_{u,t}^{(k)},\gG_{i,t}^{(k-1)};\Phi)=\text{MLP}([\rve_{u,t};\rve_{i,t};\hat{\rvx}_{u,t};\hat{\rvx}_{i,t}])
\end{equation}
where $\text{MLP}(\cdot)$ represents the MLP layer.

Given the real label $y\in\{0,1\}$ and predicted probability $\hat{y}\in\{0,1\}$, the cross-entropy loss function is adopted, formulated as:
\begin{equation}
    \gL = -\sum_{(u,i,t,y)\in\gD}[y\log\hat{y}+(1-y)\log(1-\hat{y})]
\label{eq_loss}
\end{equation}
where $\gD$ is the set of training samples.

The algorithm procedure are presented in Algorithm 1.

%% file: 4_experiment.tex
In this section, we first describe the experimental settings. Then we compare DSGL with the state-of-the-art methods. Besides, we conduct ablation study on DSGL to justify the essential components .
\input{00_res_overall}
\subsection{Experimental Setups}
\subsubsection{Datasets}
We evaluate our methods on two real-world public datasets, i.e., Amazon Dataset\footnote{http://snap.stanford.edu/data/amazon/productGraph/} and Alimama Dataset\footnote{https://tianchi.aliyun.com/dataset/dataDetail?dataId=56}. 
Amazon Clothing Dataset contains 2.8 million logs with 39 hundred users and 23 hundred items and each log contains three fields, i.e., user ID, item ID, and category. We perform a temporal train-test split, i.e., dividing the dataset into training and testing dataset by a cut timestamp, with the first 85\% for training and the rest 15\% for testing.
Alimama Dataset contains 26 million logs with 1.14 million users and 0.84 million items, and each log is composed of 14 feature fields including user ID, item ID, user age level, occupation and some other information. We use the logs in the first 7 days for training, and logs in the last day for testing.
\subsubsection{Compared Methods}
The compared methods can be grouped into conventional, sequential and graph-based categories. 
\begin{itemize}[leftmargin=10pt]
    \item \textbf{Conventional methods:} 
    \begin{itemize}
        \item \textbf{PNN}~\cite{qu2016product} uses a product layer to capture high-order feature interactions between interfield categories.
    \end{itemize}
    \item \textbf{Sequential methods:} 
        \begin{itemize}
        \item \textbf{DIN}~\cite{zhou2018deep} uses the attention mechanism to activate related user behaviors.
        \item \textbf{DIEN}~\cite{zhou2019deep} uses GRU with attentional update gate to model users' dynamic interest that is relative to the candidate item.
        \item \textbf{TIEN}~\cite{li2020deep} leverages GRU with attention mechanism to capture both time-aware user behaviors and item behaviors.
    \end{itemize}
    \item \textbf{Graph-based methods:} 
    \begin{itemize}
        \item \textbf{NGCF}~\cite{wang2019neural} is a GCN-based collaborative filtering method. It explicitly integrates a bipartite graph structure into the embedding learning process to model the high-order connectivity.
        \item \textbf{LightGCN}~\cite{he2020lightgcn} simplifies the design of GCN to make it appropriate for recommendation by light graph convolution and layer combination.
    \end{itemize}
\end{itemize}
\subsubsection{Metrics}
We adopt two widely used metrics for the CTR prediction task, i.e., \textbf{AUC} and \textbf{LogLoss}. AUC (the area under the ROC curve) measures the probability that a random clicked sample is ranked higher than a random non-clicked sample. LogLoss is the cross-entropy loss on the test dataset.

\subsubsection{Reproducibility}
For PNN, DIN, DIEN and TIEN, we use the the open-source implementations\footnote{https://github.com/itemevolutionnet/ItemEvolutionNet}. For NGCF\footnote{https://github.com/xiangwang1223/neural\_graph\_collaborative\_filtering} and LightGCN\footnote{https://github.com/kuandeng/LightGCN}, we use the source code provided by the authors.
Further implementation details and the codes are provided in the supplementary material.
\subsection{Performance Comparison}

To demonstrate the overall performance of the proposed model, we compare DSGL with the state-of-the-art recommendation methods. All experiments are repeated 10 times and averaged results are reported in Table~\ref{tab:overall}\footnote{Since LightGCN and NGCF load the whole graph in memory, causing memory overflowing on large-scale graphs with million nodes, so we didn't report the performance on Alimama Dataset.}. We have the following observations.

\begin{itemize}[leftmargin=*]
\item DSGL consistently outperforms all other baselines on both datasets. The improvements on AUC scores of DSGL over the best baseline model are 1.51\% and 3.24\%. In our practice, even 1\% improvement in AUC is substantial to achieve significant online promotion.
\item The performance of the static-graph-based methods, i.e., LightGCN and NGCF, are not competitive. The reasons are two folds. First, these methods ignore the new interactions in the testing set in the inference phase. Second, since they do not model the temporal dependency of interactions, they cannot capture the evolving interests, degrading the performances compared with sequential models.
\item All of the sequential models outperform the conventional methods and static-graph-based methods by a large margin, proving the effectiveness of capturing temporal dependency in recommendation. 
\end{itemize}

\input{00_res_ablc}
\input{00_res_abtime}
\input{00_res_abatt}

\subsection{Ablation Study}
In this section, we perform the ablation studies to show the necessity of the graph structure, and verify the effectiveness of the proposed Time-Aware Sequence Encoding Layer, Target-Preference Dual Attention Layer and Layer Combination.
\subsubsection{Effectiveness of Graph Structure and Layer Combination.}
Table~\ref{tab:ab_lc} shows the results of DSGL at different layers and its variant \textbf{DSGL w/o LC} that use the last layer instead of the combined layer as the final representation.
We have the following observations:
\begin{itemize}[leftmargin=*]
    \item Focusing on \emph{DSGL} with layer combination, the performance gradually improves with the increasing of layers. We attribute the improvement to the collaborative information carried by the multi-hop connectivity in the graph structure.
    \item Comparing \emph{DSGL} and \emph{DSGL w/o LC}, we find that removing the layer combination degrades the performance largely, which demonstrates the effectiveness of layer combination.
\end{itemize}

\subsubsection{Effectiveness of the Time-Aware Sequence Encoding Layer.}
In DSGL, we perform the Time-Aware Sequence Encoding Layer to preserve both the temporal dependency of behaviors and the fine-grained time information. 
Thus, we design ablation experiments to study how the temporal dependency and time information in DSGL contributes to the final performance. To evaluate the role of time information, we test the removal of time feature (i.e., \textbf{w/o time}). 
To evaluate the contribution of the behavior order, we test the removal of the sequence encoding module while retaining time information (i.e., \textbf{w/o Seq ENC}) and the removal of the Time-Aware Sequence Encoding Layer (i.e., \textbf{w/o TASE}). The comparison is shown in Table~\ref{tab:ab_time}. We have the following observations:

\begin{itemize}[leftmargin=*]
    \item DSGL outperforms \emph{DSGL w/o TASE} by a significant margin, demonstrating the efficacy of the Time-Aware Sequence Encoding Layer.
    \item Comparing \emph{DSGL w/o time} with the default DSGL, we observe that removing the fine-grained time decay information will cause performance degradation.
    \item DSGL outperforms \emph{DSGL w/o Seq ENC}, confirming the importance of temporal dependency carried by the historical behavior sequence.  
\end{itemize}

\subsubsection{Effectiveness of the Target-Preference Dual Attention Layer.}
In DSGL, we propose a Target-Preference Dual Attention Layer that consists of the target-aware attention and the preference-aware attention to eliminate noise from unreliable neighbors. To justify its rationality, we explore different choices here. We test the performance without the proposed attention (i.e., \textbf{DSGL w/o ATT}). We also remove the target-aware attention part (i.e., \textbf{DSGL w/o TAATT}) and the preference-aware attention part (i.e., \textbf{DSGL w/o PAATT}) from the dual attention layer respectively. From the results in Table~\ref{tab:ab_att}, we have the following observations:
\begin{itemize}[leftmargin=*]
    \item The best setting in all cases is adopting the Target-Preference Dual Attention (i.e., the current design of DSGL). Removing either the target-aware part or the preference-aware part drops the performance, demonstrating the effectiveness of dual attention in activating related neighbors and eliminating the noise.
    \item When the attention mechanism (i.e., \emph{DSGL w/o ATT}) is removed, the performance degrades largely. In some cases, the performance is even not as good as the best baseline. The observation demonstrates the necessity to introduce the attention mechanism in GNN-based recommendation methods due to the inevitable noise in the multi-hop neighborhood.
\end{itemize}

%% file: 00_res_overall.tex
\begin{table}[]
\centering
\setlength\abovecaptionskip{2pt}
\caption{Overall Comparison. The bold value marks the best one in one column, while the underlined value corresponds to the best one among all baselines.}
\label{tab:overall}
\begin{tabular}{c|cc|cc}
\toprule
\multirow{2}{*}{Method} & \multicolumn{2}{c|}{Clothing}  & \multicolumn{2}{c}{Alimama}                \\ \cline{2-5} 
                        & Logloss $\downarrow$        & AUC $\uparrow$          & Logloss $\downarrow$              & AUC $\uparrow$                                                 \\ \hline
PNN                     & 0.5817         & 0.7606                     &    0.2016            & 0.6003                     \\ \hline
DIN                     & 0.5723         & 0.7710                      &  \underline{ 0.1996  }          & 0.6116 \\ 
DIEN                     &  0.5724        &     0.7702             &      0.2006         &    \underline{0.6138}           \\ 
TIEN                    & \underline{0.5214}         & \underline{0.8166}                  &     0.2004           &   0.6125              \\ \hline
NGCF                    &  0.6734        &     0.6286                 &  -          & 	  -            \\ 
LightGCN                &   0.6578             &   0.6582             & - & -                     \\ \hline
DSGL                    & \textbf{0.5063}         & \textbf{0.8289}                     & \textbf{0.1976}               & \textbf{0.6337}             
\\ \bottomrule
\end{tabular}
\end{table}

%% file: 00_res_ablc.tex
\begin{table}[]
\centering
\setlength\abovecaptionskip{2pt}
\caption{Results of DSGL at different layers and the variant that does not use layer combination (i.e., w/o LC).}
\label{tab:ab_lc}
\resizebox{\linewidth}{!}{
\begin{tabular}{c|cc|cc}
\toprule
\multirow{2}{*}{Method} & \multicolumn{2}{c|}{Clothing} &  \multicolumn{2}{c}{Alimama} \\ \cline{2-5} 
                        & Logloss $\downarrow$        & AUC $\uparrow$                & Logloss $\downarrow$       & AUC $\uparrow$          \\ \hline
Layer 1            & 0.5238         & 0.8151         &      0.2027        &          0.6025            \\ 
Layer 2            & 0.5146         & 0.8240         &      0.2061        &            0.6060                 \\ 

Layer 3            & \textbf{0.5063}         & \textbf{0.8289}                      & \textbf{0.1976}               & \textbf{0.6337}                 \\ 
 Layer 3 w/o LC    & 0.5216         & 0.8167          &         0.1989      &         0.6291                     \\
\bottomrule
\end{tabular}
}
\end{table}

%% file: 00_res_abtime.tex
\begin{table}[]
\centering
\setlength\abovecaptionskip{2pt}
\caption{Performance of DSGL with different use of time information.}
\label{tab:ab_time}
\resizebox{\linewidth}{!}{
\begin{tabular}{l|cc|cc}
\toprule
\multirow{2}{*}{Method}         & \multicolumn{2}{c|}{Clothing} &  \multicolumn{2}{c}{Alimama}  \\ \cline{2-5} 
                                &  Logloss $\downarrow$       & AUC $\uparrow$         & Logloss $\downarrow$       & AUC $\uparrow$              \\ \hline
w/o time          &        0.5484 & 	0.7947  & 0.1977 & 	  0.6287        \\
w/o Seq ENC                            &    0.5148	 & 0.8223      &    0.1981 	 &       0.6246               \\ 
w/o TASE         &                0.5572	& 0.7857  &	0.1989 &	    0.6193     \\ 
DSGL                     & \textbf{0.5063}                        & \textbf{0.8289}                                 & \textbf{0.1976}               & \textbf{0.6337}                                    \\ \bottomrule
\end{tabular}
}
\end{table}

%% file: 00_res_abatt.tex
\begin{table}[]
\centering
\setlength\abovecaptionskip{2pt}
\caption{Performance of DSGL with different settings of attention.}
\label{tab:ab_att}
\resizebox{\linewidth}{!}{
\begin{tabular}{c|cc|cc}
\toprule
\multirow{2}{*}{Method} & \multicolumn{2}{c|}{Clothing} &  \multicolumn{2}{c}{Alimama}  \\ \cline{2-5} 
                        & Logloss $\downarrow$        & AUC $\uparrow$          & Logloss $\downarrow$       & AUC $\uparrow$                           \\ \hline
w/o ATT             & 0.5212                        & 0.8181                             &            0.2081        &      0.6216                                 \\
w/o TAATT                 & 0.5127                        & 0.8243                             &  0.2070            &    0.6273                     \\
w/o PAATT                 & 0.5187                        & 0.8196                             &  0.2074            &    0.6255                     \\
DSGL                     & \textbf{0.5063}                        & \textbf{0.8289}                         & \textbf{0.1976}               & \textbf{0.6337}                      \\ \bottomrule
\end{tabular}
}
\end{table}

%% file: 5_con.tex
In this paper, we focus on explicitly incorporating multi-hop collaborative signal into the CTR prediction model while capturing the dynamic evolution. We propose a novel graph-based method, named Dynamic Sequential Graph Learning (DSGL), to enhance users or items' representations by performing graph convolution over their dynamic sequential graphs.
Comprehensive experiments demonstrate that DSGL can consistently outperform the other state-of-art methods. 

Future directions include modeling long-term dependencies on the dynamic sequential graph as well as sampling reliable neighbors to eliminate noise from the source.